Chapter  # - will be assigned by editors

# THE LAVA MODEL: LEARNING ANALYTICS MEETS VISUAL ANALYTICS

Mohamed Amine Chatti, Arham Muslim, Manpriya Guliani, Mouadh Guesmi (Social Computing, Department of Computer Science and Applied Cognitive Science, University of Duisburg-Essen, Germany)

Abstract:   Human-Centered learning analytics (HCLA) is an approach that emphasizes the human factors in learning analytics and truly meets user needs. User involvement in all stages of the design, analysis, and evaluation of learning analytics is the key to increase value and drive forward the acceptance and adoption of learning analytics. Visual analytics is a multidisciplinary data science research field that follows a human-centered approach and thus has the potential to foster the acceptance of learning analytics. Although various domains have already made use of visual analytics, it has not been considered much with respect to learning analytics. This paper explores the benefits of incorporating visual analytics concepts into the learning analytics process by (a) proposing the Learning Analytics and Visual Analytics (LAVA) model as enhancement of the learning analytics process with human in the loop, (b) applying the LAVA model in the Open Learning Analytics Platform (OpenLAP) to support human-centered indicator design, and (c) evaluating how blending learning analytics and visual analytics can enhance the acceptance and adoption of learning analytics, based on the technology acceptance model (TAM).

## 1. INTRODUCTION

Despite the great enthusiasm currently surrounding the field of Learning Analytics (LA), we are still lacking evidence that LA has any obvious impact on learning (Ferguson & Clow, 2017; Gašević et al., 2015). This hinders the acceptance and adoption of LA at scale in schools, universities, and workplaces. Along with technical research questions, there are more crucial pedagogical and methodological problem areas related to the design, deployment, and evaluation of LA. These include the lack of attention to the LA cycle; limited attention to validity, reliability, and generalizability; limited attention to ethics; and little evaluation of commercially available tools (Ferguson & Clow, 2017). But, the most important reason is that most LA solutions are not adopted by the end users because they are not well aligned with user needs. The solution - which has been lacking in the LA community until now - is to follow a human-centered LA (HCLA) approach that emphasizes the human factors in LA and truly meets user needs. Having the human in the loop is the key to increase value and drive forward the acceptance and adoption of LA (Chatti & Muslim, 2019).

Following a human in the loop approach, Visual Analytics (VA) – a data science research field that has lately been growing very rapidly – can play a significant role to support the acceptance of LA. VA refers to analytical reasoning facilitated by interactive visual interfaces and aims at making data and information processing transparent. It integrates the analytic capabilities of the computer and the abilities of the human analyst, thus allowing novel discoveries and empowering individuals to take control of the analytical process (Thomas & Cook, 2005; Keim et al., 2006).

Visualization has been widely considered as a crucial step in the LA process and a variety of dashboards and indicators were proposed in the LA literature (Bakharia & Dawson, 2011; Bodily et al., 2018; Gašević et al., 2015; Jovanovic et al., 2008; Leony et al., 2012; Verbert et al., 2013, 2014). These dashboards, although they employ some visualizations for representing data, they are predominantly static and in general, afford very little interaction. Ritsos & Roberts (2014) point out that VA can transform LA to go beyond mere analysis (confirmatory analysis) to gaining various insights from the data (exploratory analysis) with the aim of meeting the objectives of self-assessment, performance determination, awareness, and adaptation. However, the application of VA is still under-investigated in current LA research and practice.

The main focus of this work is to explore how blending LA and VA can achieve an effective HCLA approach and improve the acceptance of LA. To get at this, we present and discuss the Learning Analytics and Visual Analytics (LAVA) model as a conceptual framework through which VA can be seamlessly integrated into the LA process. As a proof of concept, we apply the LAVA model in the Open Learning Analytics Platform (OpenLAP) that collects learning activities data from multiple sources and allows different stakeholders of LA to dynamically generate custom indicators that meet their needs. Furthermore, we evaluate OpenLAP in terms of usefulness and usability based on the technology acceptance model (TAM).

## 2. HUMAN-CENTERED LEARNING ANALYTICS

Learning Analytics (LA) focuses on the development of methods for analyzing and detecting patterns within this data and leverages those methods to support the learning experience. Chatti et al. (2012, 2014) propose a reference model for LA that provides a systematic overview on LA and fosters a common understanding of the key components of the LA ecosystem, based on four dimensions of the LA reference model, namely:
- What? What kind of data does the system gather, manage, and use for the analysis?
- Why? Why does the system analyze the collected data?
- Who? Who is targeted by the analysis?
- How? How does the system perform the analysis of the collected data?

In the ideal case, LA is a cyclical movement from data to analysis to action to learning (Chatti et al., 2014; Clow, 2012). LA is an iterative process generally carried out in six major stages, namely *Learning Activities*, *Data Collection*, *Data Storage & Processing*, *Analysis*, *Visualization*, and *Action* (see Figure 1). These steps are iterated, with each cycle yielding more effective learning activities.

User involvement in all stages of the LA process is the key to a wider acceptance and adoption of LA. It is vital to engage the various LA stakeholders (learners, teachers, institutions, researchers, developers, etc.) in the LA process. Especially, the learner should play an active role in the LA process, if LA tools are to serve the intended objective of improving learning (Chatti & Muslim, 2019). This is at the heart of HCLA. But how can an HCLA approach be implemented in practice? Visual analytics can help to face this challenge.

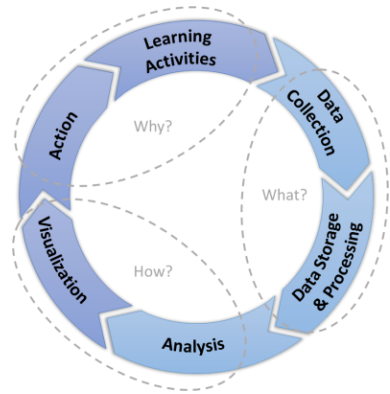

Figure 1. The learning analytics process (Chatti & Muslim, 2019)

## 3. VISUAL ANALYTICS

Computers have enormous storage capacity and computational power to support automated data analysis processes. By visualizing data, some patterns emerge which might not be noticeable in the raw form. Humans, on the other hand, have creativity, flexibility, background knowledge, sense of intuition and the skills that help to extract meaningful insights from data. Interaction is the glue that binds analytics and visualization with the human analysts (Endert et al., 2014). Visual Analytics (VA), a highly interdisciplinary research field, combines automated analysis techniques with human-interactive visualizations derived from a large amount of data, for effective understanding, reasoning, and decision making (Keim et al., 2009). The concept of VA can be better understood based on the sense-making loop for visual analytics, shown in Figure 2. A visualization is presented to the user based on the initial analysis of the dataset. The process then enters a loop, where the user drives the analysis process to draw new insights and accumulate knowledge through exploration. The user can interact with the visual representations to get a better understanding of the data through the different possible views or to eventually confirm hypotheses generated from previous iterations of analysis and interactions (Keim et al., 2008).

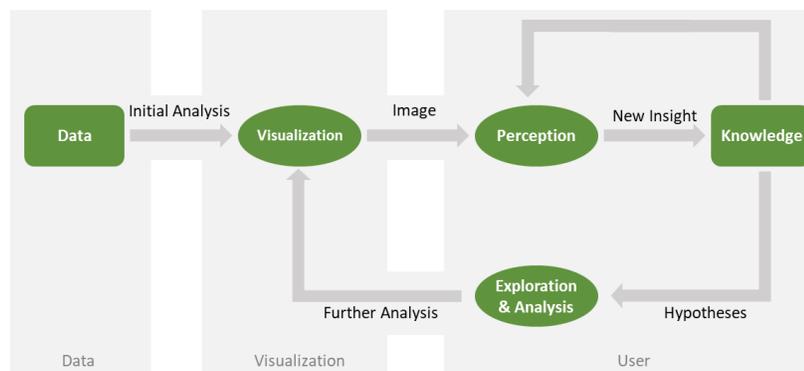

Figure 2. The sense-making loop for visual analytics (adapted from Keim et al., 2008)

## 4. THE LAVA MODEL

The main aim of LA is to turn educational data into insights, decisions, and actions in order to improve learning and teaching. However, in current LA implementations, human stakeholders are not actively involved in the LA process with its six stages namely, learning activities, data collection, data storage & processing, analysis, visualization, and action. In order to achieve HCLA, there is a crucial need to involve humans throughout the whole LA process. This is where VA comes into play. Following a human in the loop approach, VA brings humans in the data analytics process to turn data into value. This paper proposes the Learning Analytics and Visual Analytics (LAVA) model, which incorporates VA into LA and enables stakeholders to control the LA process, making it human-centered through exploratory data analysis and visualization, as depicted in Figure 3.

The LAVA model has been created by interweaving the reference model and the process of LA with the sense-making loop for VA. In this model, the four dimensions of LA (What?, Why?, Who?, How?) are revisited with VA concepts in the picture. The process of LA essentially remains the same but it is enhanced by incorporating the human perspective in the Who? dimension and exploration in the How? dimension. In the following, we discuss the eight stages of the LAVA model.

- *Learning Activities*: The LAVA model starts with concrete learning activities that can occur in different learning environments and generate a large amount of educational data.
- *Data Collection*: Collecting educational data is the foundation of the LAVA model. Today we have broad access to high-volume data from a variety of sources. The data can come from multiple, fragmented, often heterogeneous, formal, as well as informal, learning channels. It can also come in different formats, distributed across space, time, and media.
- *Data Storage and Pre-processing*: The collected data needs to be systematically stored and processed to prepare data for further analysis.
- *Analysis*: After the data is collected and pre-processed, an initial analysis is performed before presenting any results to the users.
- *Visualization*: The analysis results are presented as indicators to the users in the form of visualizations that can help to understand and interpret the results as well as to infer conclusions from the data, which can improve the learning process. These visualizations should be interactive in nature allowing users to better understand the underlying data and analyze it further.
- *Perception and Knowledge*: Users play an important role in the LAVA model. Instead of automated analysis, the users drive the entire analysis process and since every user is unique, there is no fixed path for this process. Users perceive a visualization, based on their previously acquired knowledge, the tasks which need to be done, or the goal which is intended to be achieved. Different users might draw different insights from the exact same visualization. These insights may be completely new and augment the users' knowledge base. Based on these insights, the users can generate various hypotheses which require further exploration or taking actions.
- *Exploration*: The derived insights may not always be the end results. Users might also come up with further questions. The process then enters a loop, where users steer the analysis process to draw new insights through exploration, either by directly interacting with the visualizations or by modifying different parameters of the analysis. This process can go on for as long as required until the users find answers to prove or disprove their hypotheses.
- *Action*: Taking actions is the primary aim of the LAVA model. Based on the gained knowledge, different actions, such as intervention, feedback, and recommendation may be performed with the aim of improving the learning activities.

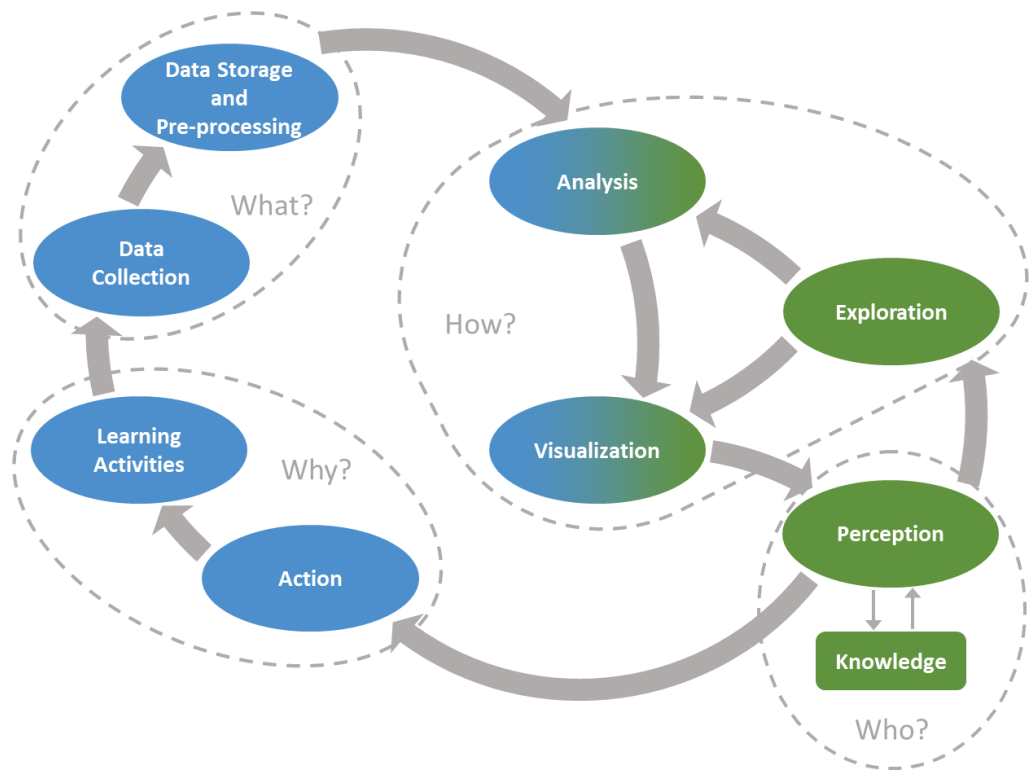

Figure 3. The LAVA model - having the human in the loop

## 5. THE LAVA MODEL IN ACTION

As a proof of concept, we applied the LAVA model in the Open Learning Analytics Platform (OpenLAP). OpenLAP is an ecosystem for Open Learning Analytics (OLA). The primary aim of OpenLAP is to collect heterogeneous data from multiple sources and support end users in defining custom indicators that meet their needs. An abstract architecture of OpenLAP is shown in Figure 4. Based on the LAVA model, the 'Indicator Editor' which is a component in the "Indicator Engine" of OpenLAP provides non-technical end-users an intuitive and exploratory user interface (UI) that gives them control in defining their own indicators in a simple way. Following the Goal - Question - Indicator (GQI) approach proposed by Muslim et al. (2017), the 'Indicator Editor' supports end users in a continuous LA process by setting appropriate analytics goal, formulating LA question, and defining indicators to answer the question. For each indicator, the user interacts with the 'Indicator Editor' to explore the stored learning activity data, apply various data filters, specify an analytics method to analyze the data, and select an appropriate visualization technique to visualize the indicator. After finalizing the indicators, they are saved and the HTML and JavaScript based indicator request codes (IRC) are generated which can be embedded in any client application to provide analytics in context (Chatti et al, 2017; Muslim et al., 2016, 2018).

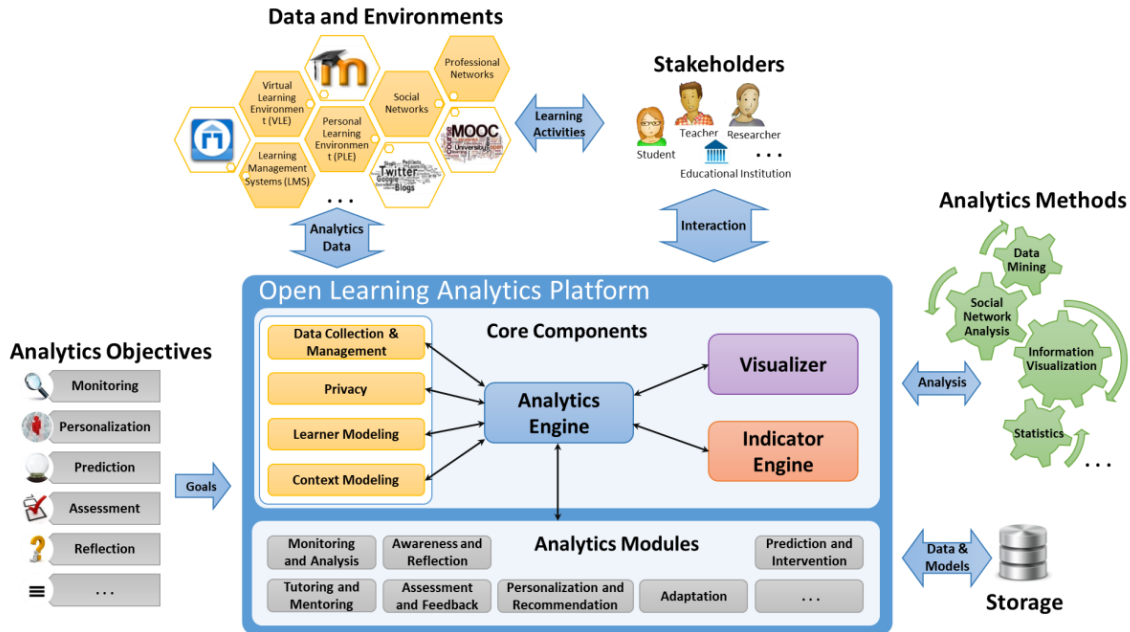

Figure 4. The abstract architecture of OpenLAP (adapted from Chatti et al., 2017)

The 'Indicator Editor' supports three different types of indicators, namely basic, composite, and multi-level analysis. The conceptual flow of each indicator type is shown in Figure 5. The basic indicator is a simple indicator type that is precisely mapped to the GQI approach. The user generates a new indicator by defining a dataset, applying various filters, selecting an analytics method for analysis, and specifying the visualization technique to render the indicator. Using this type, simple statistical indicators can be easily generated, such as "activities of discussion forum per week" and "distribution of points in assignments". The composite indicator type allows the user to combine multiple basic indicators to form a composite indicator. Using this type, indicators like "my assignment points compared to an average of my peers" and "my learning resources views compared to the average of others" can be generated. The main condition for this type is that all the basic indicators to be combined should apply the same analytics method, whereas, the dataset and filters can be different. The analysis results from each basic indicator are combined to provide a cumulative analyzed data which is rendered using the specified visualization technique. The multi-level analysis indicator type is used to generate complex indicators beyond simple statistical ones. These include indicators based on social network analysis or data mining methods, such as "cluster of students based on their learning resources views and average assignment points" and "predict students' success rate". Similar to the composite indicator type, the multi-level analysis indicator allows the user to define multiple basic indicators to define the first-level analysis. However, the user does not have to apply the same analytics methods for the basic indicators. Then, the user has to specify an attribute common in all the basic indicators based on which the analyzed data is merged and passed on to the second-level analysis. Finally, the result of the second-level analysis is rendered using the selected visualization technique. Concrete examples of composite indicators and multi-level analysis indicators are discussed in Section 5.7.3. In the following, we discuss the implementation of the 'Indicator Editor' in terms of the phases of the LAVA model.

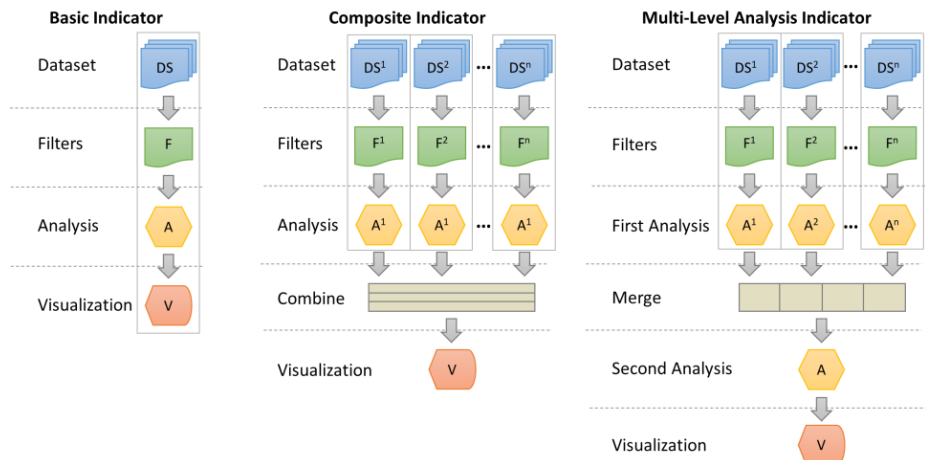

Figure 5. Indicator types supported in the OpenLAP Indicator Editor

## 5.1 Learning Activities

Learners generate a tremendous amount of interaction data while performing various activities in different learning environments. These activities include reading, writing, accessing and uploading learning material, taking tests, watching videos, and collaborating in wikis and discussion forums. Learners and teachers can use the 'Indicator Editor' to get actionable insights from this interaction data.

## 5.2 Data Collection

OpenLAP provides mechanisms to collect interaction data from different learning environments. For each source, a data collection component (collector) needs to be developed. It can either be an integrated component in a source that gathers data and pushes it to OpenLAP or an intermediate component (adapter) that receives data from a source and transforms it into a required data format before sending it to OpenLAP.

## 5.3 Data Storage and Pre-Processing

OpenLAP processes the heterogeneous data coming from different sources and stores it in the data model called Learning Context Data Model (LCDM) proposed by Thüs et al. (2014). LCDM represents a user-centric, modular, and easy to understand data model that holds additional semantic information about the context in which a learning activity has happened (Lukarov, et al., 2014). The data storage mechanism in OpenLAP is flexible and can easily be modified to make OpenLAP work with other data models, such as xAPI and IMS Caliper (Muslim et al., 2016, 2017).

## 5.4 Analysis

OpenLAP adopts a modular and extensible architecture that allows the easy integration of new analytics methods, ranging from basic statistical methods to more advanced methods like clustering, classification, and social network analysis (Muslim et al., 2018).

## 5.5  Visualization

OpenLAP allows easy integration of new visualization techniques due to its modular and extensible architecture (Muslim et al., 2018). Currently, different visualization types are available in OpenLAP based on the 'Google charts' and 'C3.js' visualization libraries, such as bar, pie, line, box plot, scatterplot. The 'Indicator Editor' allows users to try out different visualizations during the indicator generation process and select the appropriate visualization according to their needs. Figure 6 shows sample visualization generated with the 'Indicator Editor'.

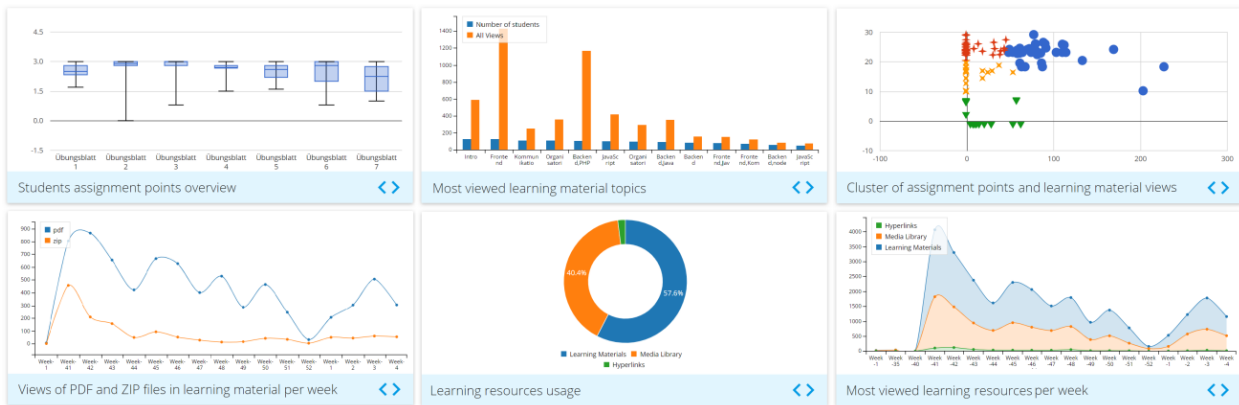

Figure 6. Sample visualizations generated with the OpenLAP Indicator Editor

## 5.6  Perception and Knowledge

Selecting appropriate visualizations in the 'Indicator Editor' can effectively help users find patterns and get actionable insights into the data. Users perceive a visualization, based on their previously acquired knowledge. Based on the visualization, users can generate various hypotheses. The process then enters a loop, where the users drive the analysis process to accumulate new knowledge through exploration in order to prove or disprove their hypotheses. Newly drawn insights can add to the knowledge of the users.

## 5.7  Exploration

The 'Indicator Editor' in OpenLAP is an interactive conversational component that enables users to control the indicator generation process, according to their needs. Following the GQI approach, the 'Indicator Editor' allows end users to generate indicators by setting analytics goal, formulating questions, and defining indicators to answer these questions. The 'Indicator Editor' supports users in exploring, analyzing, and visualizing the data through dynamic interactions, including filtering the data used for the indicator, selecting the analytics methods and visualization techniques, and changing the parameters of the data analysis algorithms. In the following, we present in detail the different sections of the 'Indicator Editor' using an example of an instructor who wants to monitor the activities of her students in a specific course.

### 5.7.1  Goal

The first step in the indicator generation process is to select an appropriate LA goal, such as assessment, intervention, monitoring, prediction, or recommendation, as shown in Figure 7. A predefined list of LA goals and their descriptions is provided to the users to help them in selecting an appropriate goal. However, if the provided list does not contain the required LA goal, the user has an option to request a new one, which will be reviewed by OpenLAP administrators and then added to the list. In our example of the instructor who wants to monitor the activities of her students, the 'Monitoring' is selected as an appropriate LA goal.

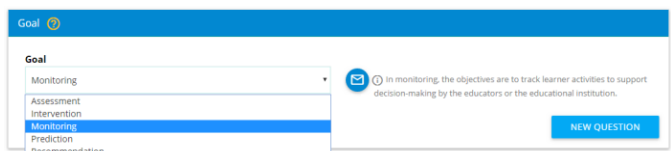

Figure 7. The Goal section in the OpenLAP Indicator Editor

### 5.7.2 Question

After selecting an appropriate LA goal, the next step in the indicator generation process is to formulate a suitable LA question. Afterwards, multiple indicators can be associated with the LA question either by loading an existing indicator generated by another user and modifying its parameters or defining a new basic, composite or multi-level analysis indicator, as discussed in the next section. While a question might be asked in an abstract manner, an indicator is a concrete calculator with a corresponding visualization to answer the question. In our example, the instructor entered "How active are students in my class?" as the LA question and defined a set of four indicators to answer this question, namely Students weekly learning resources access", "Students assignment points overview", "Most viewed learning materials", and "Correlation of assignment points and learning resources views", as shown in Figure 8. She can then view, delete, or select any associated indicator for editing again. Finally, she can visualize the question and the associated indicators in a dashboard format and save them to get the indicator request code (IRC) for the whole question as well as for each individual indicator, as shown in Figure 9. The IRC is composed of HTML and JavaScript code which can be embedded in any client application that allows web content (e.g. dashboards, Web pages).

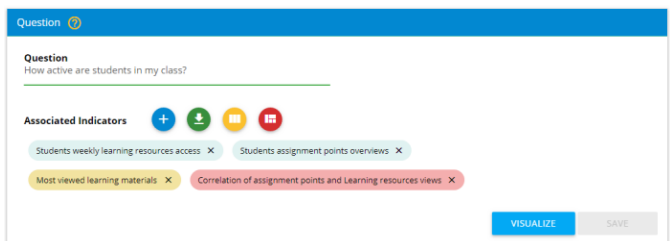

Figure 8. The Question section in the OpenLAP Indicator Editor

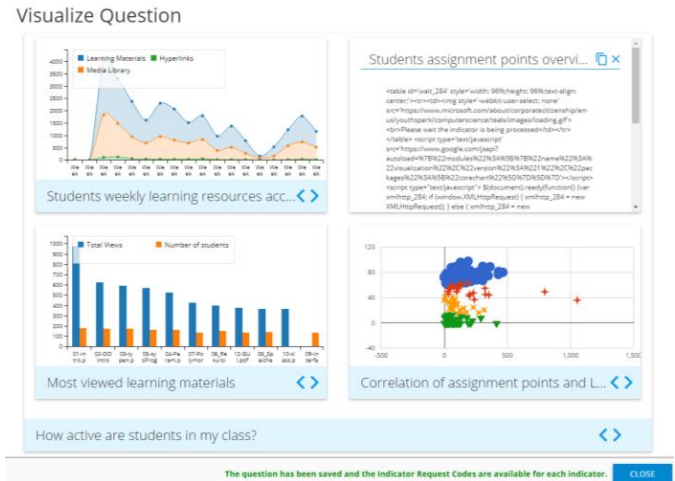

Figure 9. The Visualize Question section in the OpenLAP Indicator Editor containing Indicator Request Code (IRC)

### 5.7.3 Indicator

Three different indicator types, namely basic indicator, composite indicator, and multi-level analysis indicator can be associated with an LA question.

*Basic Indicator*
In our example, the instructor associated her question with two basic indicators "Students weekly learning resources access" and "Students assignment points overview", which are shown in blue (see Figure 8). The process of defining a basic indicator consists of four main parts, namely dataset, filters, analysis, and visualization.

- *Dataset*: This part of the 'Indicator Editor' allows the user to define the parameters for the indicator dataset. These include the list of data sources (e.g. Moodle, edX), platform types (e.g. web, mobile), performed actions (e.g. add, view, update, delete, post), and the category of objects on which the action is performed (e.g. wiki, discussion forum, assignments, learning materials). Figure 10 shows the dataset part of the 'Indicator Editor' where the required parameters have been selected for the "students weekly learning resources access" indicator.

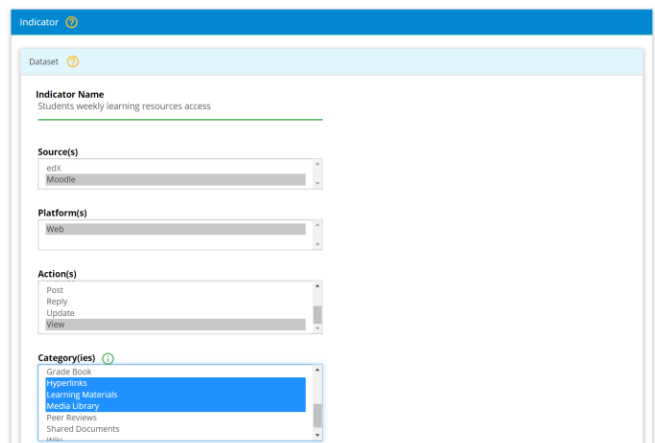

Figure 10. The Dataset part in the OpenLAP Indicator Editor to define a Basic Indicator

- *Filters*: This part allows the user to refine the selected dataset by applying various filters. Three different types of filters can be applied to the dataset, namely 'Attribute', 'Time', and 'User'. These filters are grouped under two tabs, as shown in Figure 11. The 'Attribute' filters are applied to the additional semantic information stored related to each category. E.g. for 'Learning Materials' category, attributes like 'Name', 'File Extension', and 'Size (in Bytes)' are available, whereas, for 'Assignments' category, possible available attributes are 'Title', 'Total Marks', and 'Due Date'. When multiple categories are selected, only the attributes common to all the selected categories are provided. The user can search for the possible values of the selected attribute and select one or more values. All the applied attribute filters are shown at the top, from where they can easily be removed. The 'Time / User' tab is split into the 'Time' and 'User' sections. The 'Time' filter section provides the possibility to specify the start and/or end date for which the dataset should be considered. In the 'User' section, due to the privacy concerns, the user only has the option to use the anonymized data of everyone, use own data only, or use the anonymized data of everyone excluding own data.

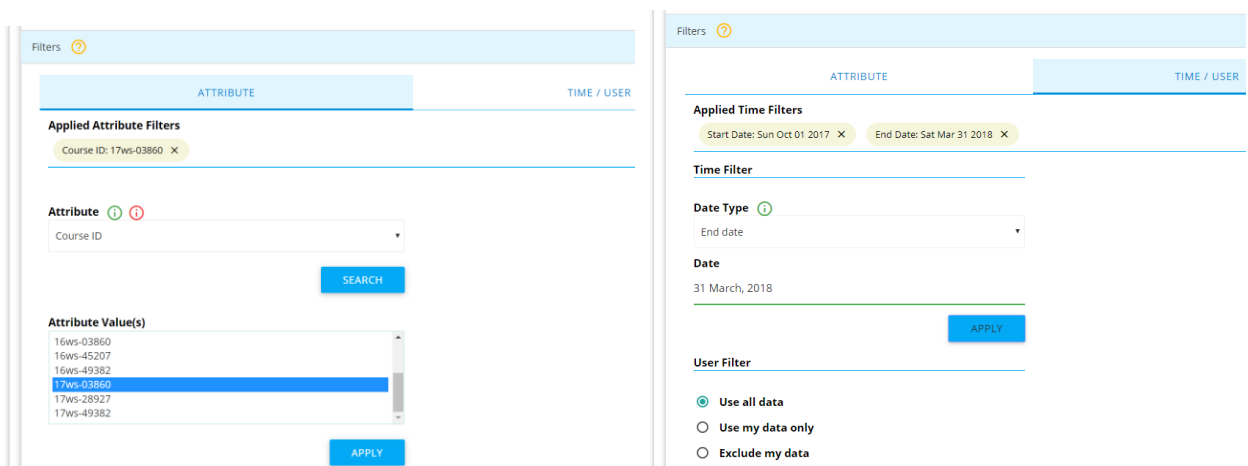

Figure 11. The Filters part in the OpenLAP Indicator Editor to define a Basic Indicator

- *Analysis*: After defining the dataset and applying the required filters, the user can specify which analytics method should be used to perform the analysis on the filtered dataset. The additional parameters required by the analytics methods are shown to the user with the default values pre-selected. The final step is to define the mappings between the filtered dataset and the selected analytics method by specifying which column of the dataset should be used for which input of the selected analytics method. The mapping is performed by selecting an analytics method input as well as the dataset column that needs to be mapped to the selected input and clicking the 'Add' button. The mapped analytics method inputs are removed from the selection and added to the list of mapped inputs. The red-colored analytics method inputs are required, and the green ones are optional. For the "students weekly learning resources access" example indicator, the instructor selected the analytics method 'Count items per week', as shown in Figure 12. This analytics method requires three inputs, namely 'Items to count', 'User' and 'Timestamp'. The '(Text)' part attached to each input indicates that the specific input can only accept the data columns that are of '(Text)' type. The 'Timestamp' input has already been mapped to the 'Timestamp' data column and it is no longer available in the analytics method inputs list.

Figure 12. The Analysis part in the OpenLAP Indicator Editor to define a Basic Indicator

- *Visualization*: The final step in the indicator generation process is to define the visualization of the indicator. In this part, the user selects an appropriate visualization library and visualization type for the indicator. Figure 13 shows the 'Stacked Area Chart' visualization type of the 'C3/D3.js' visualization library, which is used for the "students weekly learning resources access" example indicator. Similar to the analytics method part, the user defines the mapping between the outputs of the analytics method and the inputs of the selected visualization type. After defining all the parameters, the indicator can be previewed. The user can further explore by changing the dataset, applying different filters, and updating the specified mappings to come up with the indicator that fits her needs. Finally, the indicator can be associated with the LA question.

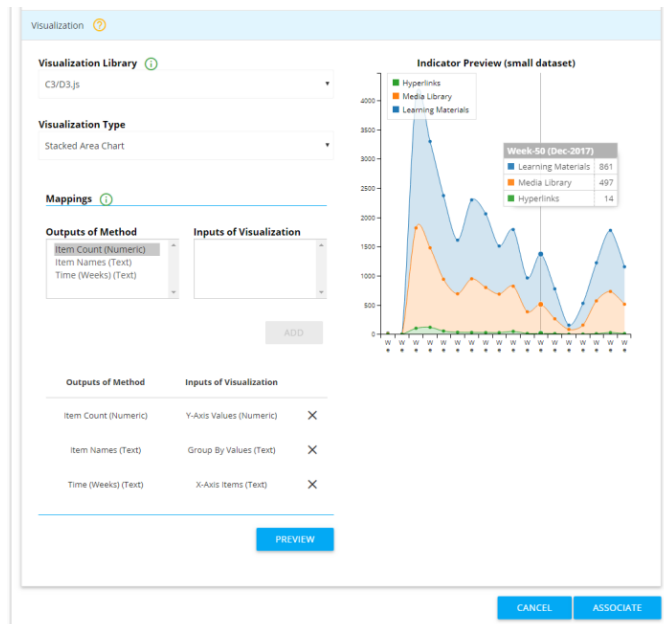

Figure 13. The Visualization part in the OpenLAP Indicator Editor to define a Basic Indicator

*Composite Indicator*
A composite indicator can be generated by combining two or more basic indicators already associated with the LA question. The main condition for the composite indicator is that all the basic indicators to be combined should apply the same analytics method. For the example LA question "How active are students in my class?", the instructor defined a composite indicator called "Most viewed learning materials" using the 'Number of students' and 'Total Views' basic indicators, as shown in Figure 14. Based on the analytics method of the first selected basic indicator, the UI notifies the user which basic indicators can be combined by highlighting them with green and disabling the others. In the example, the two basic indicators apply the analytics method 'Count N most occurring items'. The 'Number of students' indicator counts the top 10 learning materials which are viewed by the highest number of students and the 'Total View' indicator counts the top 10 most viewed learning materials. After selecting the required indicators to combine, the user selects an appropriate visualization library and visualization type to preview the indicator. Finally, the composite indicator is associated with the LA question and shown in yellow (see Figure 8).

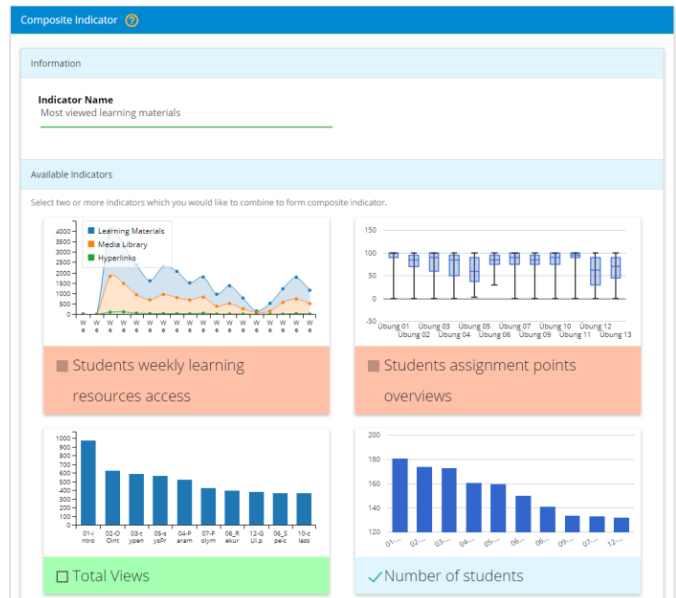

Figure 14. The Composite Indicator section in the OpenLAP Indicator Editor

*Multi-Level Analysis Indicator*:

The multi-level analysis indicator consists of three parts, namely first-level analysis, second-level analysis, and visualization. Figure 15 shows the UI for the first-level analysis part where the instructor selected two basic indicators, namely 'Views' and 'Points' to create a multi-level analysis indicator "Correlation of assignment points and learning resources views". Next, the analyzed datasets from the basic indicators are merged by selecting a common attribute and passed on to the second-level analysis to identify clusters of students based on their assignment points and learning material views. The second-level analysis and the visualization parts are similar to the analysis and the visualization parts of the basic indicator. Finally, the multi-level analysis indicator is associated with the LA question and shown in red (see Figure 8).

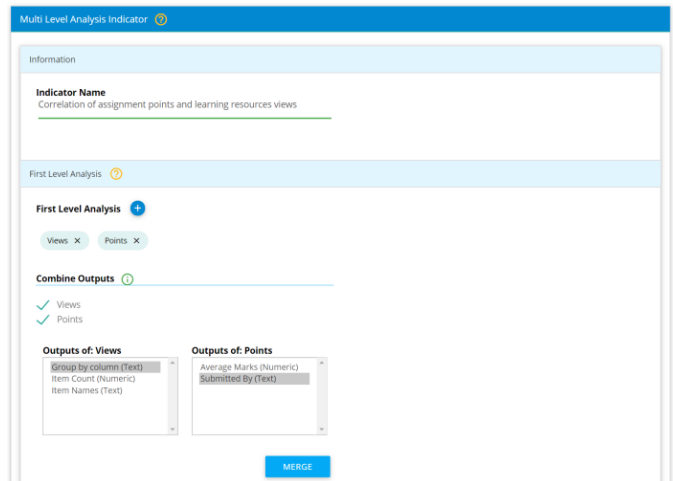

Figure 15. The Multi-Level Analysis Indicator section in the OpenLAP Indicator Editor

## 5.8 Action

The aim of the 'Indicator Editor' is to help users turn data into decisions and actions (e.g. intervention, feedback, recommendation) by creating custom indicators, following the GQI approach. Since users are steering the indicator generation process, the insights drawn from these indicators will fit their needs and thus lead to useful actions. This is at the heart of LAVA model.

## 6. EVALUATION

A thorough evaluation of the indicator generation process was conducted based on the technology acceptance model (TAM) to assess the user acceptance of the OpenLAP 'Indicator Editor'. Based on TAM, the two most important factors that influence the user's decision to use the system are perceived usefulness (PU) and perceived ease of use (PEOU) (Davis et al., 1989). Therefore, the 'Indicator Editor' is evaluated in terms of usefulness and usability.

## 6.1 Method

We employed various techniques to perform a quantitative and qualitative evaluation of the 'Indicator Editor' in terms of usefulness and usability (Dix et al., 2003). The cognitive walkthrough evaluation approach was used with the participants who were experts in the field of data analytics and LA and the think aloud method was used with the students (Nielsen, 1994). The participants were asked to use the 'Indicator Editor' and generate a set of different indicators. Moreover, a custom questionnaire was also filled out by the participants at the end of the interview sessions, which contained questions related to the overall usefulness of the 'Indicator Editor' and its usability based on the System Usability Scale (SUS) (Brooke, 1996). These questions were also discussed with the participants in the interview sessions.

### 6.1.1 Setting

OpenLAP was seamlessly integrated as a third-party system in the university LMS and made available to the participants of three courses. 414 students and 12 teachers volunteered to take part in the study. After a month of using the system, the participants were requested to take part in semi-structured interview sessions for the purpose of collecting detailed facts, opinions, and critiques about different aspects of using the 'Indicator Editor'. Each interview session lasted 30 to 60 minutes depending on the pace at which the participant performed the tasks and the amount of provided feedback. The audio, video, and screen recording for each interview session were captured. The data collected from each session was analyzed to improve the design of the questions for the next interview sessions. After all the sessions were finished, the next step was to clean, organize, and examine the results and draw conclusions. Thus, the audio recording of each interview was carefully transcribed. Additional information was extracted from the videos as well as from the screen recordings and embedded in the text. Afterwards, the data from all the interview sessions were coded based on the coding process proposed by Corbin & Strauss (1990), combined, and analyzed to generate understanding and derive patterns.

### 6.1.2 Participants

The participants who volunteered for the semi-structured interview sessions included 34 students and 5 teachers. The students were enrolled in either a Bachelor (70%) or a Master (30%) degree program in Computer Science. About 28% of the participants told that they were not familiar with data analytics concepts and about 51% were not familiar with LA concepts. About 54% and 36% of the participants mentioned that they were aware of the concepts of data analytics and LA respectively.

## 6.2 Usefulness

The questionnaire contained the following questions to gauge the overall usefulness of the 'Indicator Editor'.
- Do you think that by using the 'Indicator Editor' you can now generate the required indicators to support your learning/teaching activities?
- Do you think that the 'Indicator Editor' is a useful tool to improve your teaching/learning experience?
- Do you think that the 'Indicator Editor' provides the interaction possibilities that you expect from it?
- Are you satisfied with the level of interactivity to support indicator customization provided by the 'Indicator Editor'?

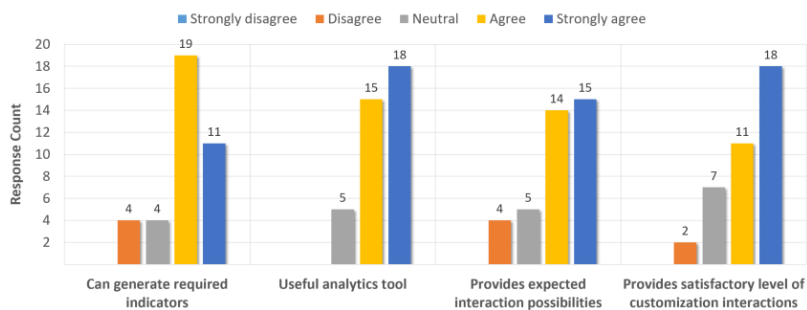

Figure 16. Overall usefulness of the OpenLAP Indicator Editor

Around 79% of the participants thought that they will be able to generate indicators that can support their learning/teaching activities, as shown in Figure 16. All the teachers said that the 'Indicator Editor' provides enough possibilities to generate the required indicators. One teacher further added that this is dependent on the quality and the amount of data available in OpenLAP. Most of the students also agreed that they can easily define the required indicators. However, they might need some time to try out different options and learn the system to be able to generate advanced indicators. 87% of the participants agreed that the 'Indicator Editor' has the potential to be a useful analytics tool to improve their teaching/learning experience. One student said that *"normally just the teachers have this information via the evaluations and now I am on this side and I can evaluate myself"*. In response to the question of the 'Indicator Editor' providing the expected interaction possibilities, 78% of the participants replied positively. Around 76% of the participants agreed that the level of interactivity has a positive effect on the indicator customization. Some participants stated that the exploration options provided in the 'Indicator Editor' would lead to customized indicators that meet their needs. However, some of the participants suggested to design two separate modes of the UI, one with detailed customization possibilities for technical users and another one with just simple options for non-technical ones.

The interview sessions focused on the usefulness of the GQI approach in the 'Indicator Editor'. In general, the GQI approach was perceived as an intuitive way to structure the process of defining custom indicators. Some participants mentioned that the effect of selecting a specific LA goal on the overall indicator generation process was not clear. However, after explaining that as future work we are planning to extend the 'Indicator Editor' with a feature that provides recommendations of questions and indicators based on the selected goal, almost everyone agreed on the importance of specifying a goal at the beginning of the indicator generation process. Some participants did not understand the relationship between the question and indicators. For example, a student said that *"the goal is abstract that I understood ... that was logical to me ... I was not sure how a question is different from an indicator"*. As an improvement, some participants suggested to provide sample questions and related indicators in the 'Indicator Editor'.

## 6.3 Usability

The usability of the 'Indicator Editor' is calculated using the System Usability Scale (SUS), which is a simple ten-item attitude Likert scale giving a global view of subjective assessments of usability (Brooke, 1996). Based on the results of the custom questionnaire, the SUS score of the 'Indicator Editor' is calculated to be approximately 61. The results of the overall usability evaluation are shown in Figure 17. 74% of the participants agreed that they would use the 'Indicator Editor' frequently, e.g. at the beginning of a semester when the courses start, to set up their personal dashboards with the required indicators or when they might need an indicator which is not already available in the indicator catalog. However, there is relatively less agreement related to the ease of use of the system. According to the participants, this is mainly due to the unfamiliarity with the indicator generation process, the mapping steps, and the usage of some terms which might not be easy to understand by non-technical users, e.g. columns, mapping, and attributes. Some participants suggested again to have beginner and advanced modes of the UI. In terms of learnability, some participants stated that the system requires some effort and time to understand and start working with it and requested to provide video tutorials of different possible scenarios in the system.

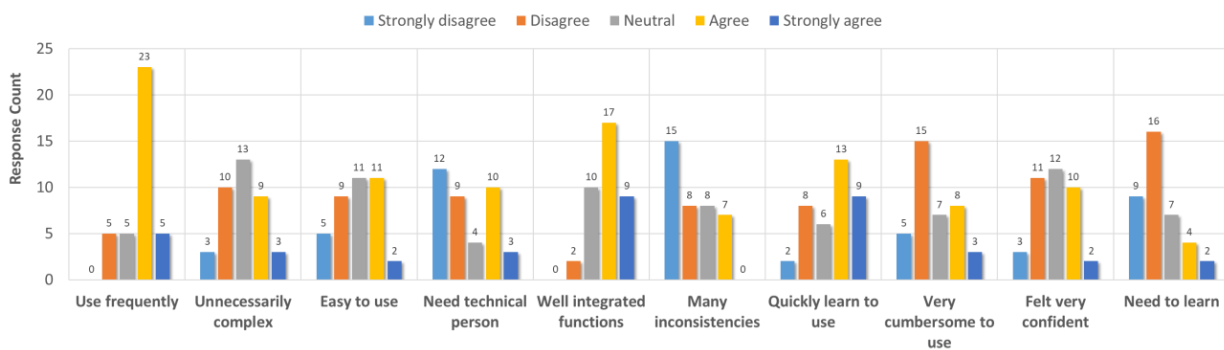

Figure 17. Overall usability of the OpenLAP Indicator Editor

The interview sessions revealed that defining mappings in the analysis and visualization parts of the indicator generation process (see Section 5.7.3) was the most complex task. For example, a student said that *"this (mapping) is cool but I think it's really hard to do this on your own ... especially for beginners"*. In order to improve the usability of the mapping steps, participants suggested to include automated mappings, e.g. *"if I have 'Timestamp (Numeric)' as input and have the same column name then it's the obvious mapping ... so at least it should be automatic and then if required I can change it"* and *"if you select something 'Text' based then you don't see 'Numeric' values in the other list ..."*. Participants also suggested to provide examples of the expected inputs/outputs, e.g. *"I would prefer to directly add ... maybe in curly brackets after dataset column name, one or two examples ... like 'Source (Text) {edX, Moodle}' ... then it would be clear"*.

## 7. CONCLUSION

In this paper, we argued that adopting a human-centered learning analytics (HCLA) approach is vital to improve the acceptance and adoption of learning analytics. We proposed the Learning Analytics and Visual Analytics (LAVA) model as a possible implementation of the HCLA approach that, by having the human in the loop, has the potential to improve the acceptance of learning analytics. As a proof of concept, we applied the LAVA model in the Open Learning Analytics Platform (OpenLAP) to support learners and teachers in defining custom indicators that meet their needs. We conducted a mixed-method evaluation of the user acceptance of the OpenLAP 'Indicator Editor' based on the technology acceptance model (TAM). The evaluation results showed that following a LAVA model-based approach has the potential to push forward the acceptance and adoption of learning analytics.